\documentclass[prd,twocolumn,showpacs,nofootinbib]{revtex4}
\usepackage{times}
\usepackage{natbib}
\usepackage{epsfig}
\usepackage{bm}
\usepackage{comment}
\usepackage{color}

\newcommand{\np}{n_g}
\newcommand{\msun}{{\it h}^{-1}M_{\odot}}

\newcommand{\beeq}{\begin{equation}}
\newcommand{\bnobs}{\bar\np}
\newcommand{\zz}{z}      
    
\newcommand{\dz}{\delta z}           
\newcommand{\dzv}{\delta z_v}
\newcommand{\dv}{\delta_m^{(v)}}
\newcommand{\ee}{e}                 
\newcommand{\bee}{\boldsymbol{\ee}}
\newcommand{\bpp}{\boldsymbol{p}}
\newcommand{\eneq}{\end{equation}}
\newcommand{\dgr}{\delta_\up{GR}}
\newcommand{\ax}{\alpha_{\chi}}     
\newcommand{\px}{\varphi_{\chi}}    
\newcommand{\dzg}{\dz_\chi}
\newcommand{\drg}{\delta\mathcal{R}}  
\newcommand{\rs}{r}
\newcommand{\HH}{\mathcal{H}}
\newcommand{\ddL}{\delta\mathcal{D}_L} 
\newcommand{\kag}{\mathcal{K}}      
\newcommand{\bear}{\begin{eqnarray}}
\newcommand{\vx}{v_{\chi}}          

\newcommand{\dnewt}{\delta_\up{Newt}}
\newcommand{\OM}{\Omega_m}
\newcommand{\cA}{\mathcal{P}}    
\newcommand{\cB}{\mathcal{R}}    
\newcommand{\enar}{\end{eqnarray}}
\newcommand{\bdobs}{\boldsymbol{\delta}_\up{GR}^\up{obs}}

\newcommand{\bb}{\boldsymbol{b}_0}         
\newcommand{\bI}{\boldsymbol{I}}    
\newcommand{\bnn}{\boldsymbol{\varepsilon}}   
\newcommand{\BB}{\boldsymbol{b}}       
\newcommand{\up}[1]{{\rm #1}}

\newcommand{\AVE}[1]{\langle#1\rangle}
\newcommand{\BT}{\boldsymbol{b}^\dagger}     
\newcommand{\Pv}{P_m}
\newcommand{\EE}{\boldsymbol{\mathcal{E}}}       
\newcommand{\mA}{\alpha}   
\newcommand{\EEI}{\boldsymbol{\mathcal{E}}^{-1}} 
\newcommand{\mB}{\beta}   
\newcommand{\mBS}{\beta^*}   

\newcommand{\mC}{\gamma}   
\newcommand{\fnl}{f_\up{NL}}
\newcommand{\gpc}{\rm Gpc}
\newcommand{\hgpc}{{h^{-1}\gpc}}
\newcommand{\hmpci}{h\up{Mpc}^{-1}}

\newcommand{\kmin}{k_\up{min}}
\newcommand{\kmax}{k_\up{max}}

\newcommand{\ccA}{c_{\cA}}
\newcommand{\ccB}{c_{\cB}}

\begin{document}

\title{Testing General Relativity on Horizon Scales and the Primordial
non-Gaussianity}

\author{Jaiyul Yoo$^{1,2}$}
\altaffiliation{jyoo@physik.uzh.ch}
\author{Nico Hamaus$^1$}
\author{Uro{\v s} Seljak$^{1,2,3,4}$}
\author{Matias Zaldarriaga$^{5}$}
\affiliation{$^1$Institute for Theoretical Physics, University of Z\"urich,
CH-8057 Z\"urich, Switzerland}
\affiliation{$^2$Lawrence Berkeley National Laboratory, University of 
California, Berkeley, CA 94720, USA}
\affiliation{$^3$Physics Department and Astronomy Department,
University of California, Berkeley, CA 94720, USA}
\affiliation{$^4$Institute for the Early Universe, Ewha Womans University, 
120-750 Seoul, South Korea}
\affiliation{$^5$School of Natural Sciences, Institute for Advanced Study, 
Einstein Drive, Princeton, NJ 08540, USA}

\begin{abstract}
The proper general relativistic description 
of the observed galaxy power spectrum is substantially different from
the standard Newtonian description on large scales, providing a unique 
opportunity to test general relativity on horizon scales. 
Using the Einstein equations, the general relativistic effects can be 
classified as two new terms that represent the velocity and the gravitational
potential, coupling to the time evolution of galaxy number density and 
Hubble parameter. Compared to the dominant density and velocity redshift-space
distortion terms, the former scales as $\HH/k$ and correlates the real and
imaginary parts of the Fourier modes, while the latter scales as
$(\HH/k)^2$, where~$k$ is the comoving wave number and $\HH$ is the
conformal Hubble parameter.
We use the recently developed methods to reduce the sampling variance and 
shot noise to show that in an all sky galaxy redshift survey at low
redshift the velocity term can be measured at
10-$\sigma$ confidence level, if one can utilize halos of mass 
$M\geq10^{10}\msun$,
while the gravitational potential term itself can only be marginally detected.
We also demonstrate that the general relativistic effect is {\it not}
degenerate with the primordial non-Gaussian signature in galaxy bias,
and the ability to detect the primordial non-Gaussianity is 
little compromised.
\end{abstract}


\maketitle

The recent discovery of the cosmic acceleration has renewed interest in
modifications 
of gravity on cosmological scales, and tests of general relativity
on horizon scales become ever more crucial in determining whether the 
cosmic acceleration is due to dark energy or the breakdown of general
relativity. However, there exists a fundamental limitation to any attempts
for testing general relativity on large scales, and decisive conclusions
remain elusive due to the cosmic variance limit set by the finite
number of measurements available to us.

In the past few decades galaxy clustering has been one of the indispensable
tools in cosmology, covering a progressively larger fraction of the sky
with increasing redshift depth. 
With the upcoming dark energy surveys this trend
will continue in the future. However, 
despite the advance in observational
frontiers, there remained a few unanswered questions in 
the theoretical description
of galaxy clustering. On horizon scales, the standard Newtonian
description naturally breaks down, and a choice of hypersurface of simultaneity
becomes an inevitable issue, demanding a fully relativistic treatment
of galaxy clustering beyond the current Newtonian description.

In recent work \cite{YOFIZA09,YOO10}, it is shown that a proper relativistic
description can be easily obtained by following the observational procedure
in constructing the galaxy fluctuation field and its statistics: We need to
model observable quantities, rather than theoretically convenient but 
unobservable quantities, usually adopted in the standard method.
While both the relativistic and the standard Newtonian
descriptions are virtually indistinguishable in the Newtonian
limit, they are substantially different on horizon scales, rendering galaxy
clustering measurements a potential probe of general relativity. 
The relativistic description of galaxy clustering includes 
two new terms that scale as velocity and gravitational potential. 
Compared to the dominant density contribution, they are suppressed by
$\HH/k$ and $(\HH/k)^2$ and become important only on large scales, where
the comoving wavevector amplitude~$k$ is comparable to the conformal 
Hubble parameter $\HH=aH$. Consequently, the
identification of these terms just by looking at the galaxy power spectrum is
hampered  because 
of sampling variance, and the general relativistic
effects unaccounted in the standard Newtonian description may result in
systematic errors less than 1-$\sigma$ for most 
of the volume available at $z\leq3$ in the standard power spectrum analysis
\cite{YOO10}.

A new multi-tracer method \cite{SELJA09} takes advantage of the fact 
that differently biased galaxies 
trace the same underlying matter distribution, and it can be
used to cancel the randomness of the matter distribution in a single 
realization of the Universe, eliminating the sampling variance limitation. 
This method has been used in \cite{MCDON09} to investigate the velocity 
effects 
of \cite{YOFIZA09,YOO10}, noting that for any given Fourier mode
the imaginary part of velocity couples to the real part of density and vice 
versa. 
Even with this novel technique, the expected detection level is low 
\cite{MCDON09}.

If sampling variance is eliminated, the dominant remaining source of error 
is shot noise, 
caused by the discrete nature of galaxies. Recently, a shot noise cancelling 
technique has been proposed \cite{SEHADE09} and investigated for detecting 
primordial non-Gaussianity \cite{HASEDE11}
in combination with the sampling variance cancelling technique.
The basis of the method is that by using halo mass dependent weights one can 
approximate a halo field 
as the dark matter field and reduce the stochasticity between them. While 
this works best when 
comparing halos to dark matter, some shot noise cancelling can also be 
achieved by comparing halos 
to each other \cite{HASEET10,HASEDE11}.
This opens up a new opportunity to probe horizon scales and
extract cosmological information with higher signal-to-noise ratio.
In this {\it Letter} we explore the possibility of using galaxy power spectrum
measurements combined with the multi-tracer and shot noise cancelling 
techniques
to test general relativity on cosmological horizon scales.
In addition, we comment on the impact of the general relativistic effects
in detecting the primordial non-Gaussian signature in galaxy bias.

{\it GR description.$-$} A full general relativistic description of galaxy
clustering is developed in \cite{YOFIZA09,YOO10} (see also
\cite{BODU11CHLE11,JESCHI11}). Here we make one modification in the adopted
linear bias ansatz.
Previously, we have adopted the simplest linear bias ansatz,
in which the galaxy number density is just a function
of the matter density $\np=F[\rho_m]$, both at the same 
spacetime \cite{FOOTNOTE}.
However, this ansatz turns out to be rather restrictive, since the time 
evolution 
of the galaxy sample is entirely driven by the evolution of the matter density
$\propto(1+z)^3$. Now we relax this assumption and provide more freedom,
while keeping the locality: We allow the galaxy number density at the same
matter density to differ
depending on its local history, i.e., $\np=F[\rho_m,t_p]$ with $t_p$ being
the proper time measured in the galaxy rest frame. Therefore, when expressed
at the observed redshift~$\zz$, the galaxy number density can be
written as
\bear
\np&=&\bnobs(\zz)\left[1+b~(\delta_m-3~\dz)-b_t~\dzv\right] \nonumber \\
&=&\bnobs(\zz)\left[1+b~\dv-\ee~\dzv\right]~,
\label{eq:bias}
\enar
where the lapse~$\dz$ in the observed redshift~$\zz$ is defined as 
$1+\zz=(1+\dz)/a$ and the expansion factor is~$a$.
The script~$v$ indicates quantities are evaluated in the dark matter comoving
gauge ($v=0$), and this gauge condition coincides with the synchronous gauge,
as presureless dark matter can freely fall in the rest frame 
(see, e.g., \cite{HWNO99WASL09}). 

The galaxy bias factor is
$b=\partial\ln\bnobs/\partial\ln\rho_m|_{t_p}$ and
the time evolution of the galaxy number density due to its local history is
$b_t=\partial\ln\bnobs/\partial\ln(1+z)|_{\rho_m}$. Therefore, 
the total time evolution of the mean galaxy number density is 
proportional to the evolution bias \cite{YOO09},
$\ee=d\ln\bnobs/d\ln(1+z)=3~b+b_t$~. This biasing scheme in Eq.~(\ref{eq:bias})
is consistent with \cite{BODU11CHLE11,BASEET11,BRCRET11,JESCHI11}, 
and our previous bias ansatz corresponds to $b_t=0$. 
Physically, the presence of long wavelength modes affects 
the local dynamics of galaxy formation by changing the local curvature
and thus the expansion rate, and these are modulated by the Laplacian of the
comoving curvature \cite{BASEET11}. 

Therefore, with this more physically motivated bias ansatz,
the general relativistic description of the observed galaxy 
fluctuation is \cite{YOFIZA09,YOO10}
\bear
\label{eq:fullGR}
\dgr&=&\left[b~\dv-e~\dzv\right]+\ax+2~\px+V+3~\dzg \nonumber \\
&+&2~{\drg\over\rs}
-H{d\over d\zz}\left({\dzg\over\HH}\right)
-5p~\ddL-2~\kag~, 
\enar
where the luminosity function slope of the source galaxy population is~$p$,
the comoving line-of-sight distance is~$\rs$, the fluctuation in the luminosity
distance is $\ddL$, the temporal and spatial metric perturbations are
$\ax$ and $\px$, the line-of-sight velocity is~$V$, and the gauge-invariant
radial displacement and lensing convergence are $\drg$ and $\kag$.
We have ignored the vector and tensor contributions to $\dgr$.
The subscript~$\chi$ indicates quantities are evaluated in 
the conformal Newtonian gauge ($\chi=0$).
We emphasize that compared to \cite{YOFIZA09,YOO10}
it is only the terms in the square bracket that are
affected by the choice of linear bias ansatz.

Noting that Eq.~(\ref{eq:fullGR}) can be most conveniently computed in the
conformal Newtonian gauge, we first remove the lapse term~$\dzv$ by
using its gauge transformation property $\dzv=\dzg+\HH\vx/k$, but we keep
the comoving gauge matter fluctuation~$\dv$ to be consistent with the
convention in the literature.
In Fourier space, a further simplification can be made by ignoring the
projected quantities such as the gravitational lensing and the integrated 
Sachs-Wolfe contributions, which are important only 
for the pure transverse modes ($k^\parallel=0$) \cite{HUGALO08,YOO10}.
Equation~(\ref{eq:fullGR})
in Fourier space is then
\begin{widetext}
\bear
&&\hspace{-10pt}
\dgr=\left[b~\dv-\mu_k^2{k\vx\over\HH}\right] -\ee~{\HH\over k}~\vx
+2~\px-{\px'\over\HH}+i\mu_k{k~\ax\over\HH}-i\mu_k{\vx'\over\HH}
+\left[\ee+1-{1+\zz\over H}{dH\over dz}+(5p-2)\left(1-{1\over\HH\rs}\right)
\right]\ax \nonumber \\
&&
+\left[\ee-1-{1+\zz\over H}{dH\over dz}+(5p-2)\left(1-{1\over\HH\rs}
\right)\right]i\mu_k\vx 
=\dnewt-i\mu_k
\left[\ee-{1+\zz\over H}{dH\over dz}+(5p-2)\left(1-{1\over\HH\rs}
\right)\right]{f~\dv\over k/\HH} \nonumber \\
&&
+\bigg\{\ee f-{3\over2}\OM(z)\bigg[\ee+f-2-{1+\zz\over H}{dH\over dz}
+(5p-2)\left(1-{1\over\HH\rs}\right)\bigg]\bigg\}{\dv\over (k/\HH)^2}
\equiv\dnewt-i\mu_k{\cB~\dv\over k/\HH}+{\cA~\dv\over (k/\HH)^2}~,
\label{eq:full}
\enar
\end{widetext}
where $f$ is the logarithmic growth rate and we defined the
two dimensionless coefficients $\cB$ and~$\cA$ in the last equality.
We have used the Einstein equations in the second equality to express 
$\ax$, $\px$, and $\vx$ in terms of $\dv$.
The standard Newtonian description of the observed galaxy 
fluctuation is $\dnewt=b~\dv-\mu_k^2~k\vx/\HH=(b+f\mu_k^2)~\dv$, which can 
be contrasted with $\dgr$ in Eq.~(\ref{eq:fullGR}).

Apparent from their spatial dependence in Eq.~(\ref{eq:full}),
the coefficients $\cB$~and~$\cA$
originate from the velocity and gravitational potential.
While $\cA$ is purely relativistic, some
contributions to~$\cB$ may be considered non-relativistic, since they 
could be written down in Newtonian dynamics,
simply as a coupling of velocity from the Doppler effect with
the time evolution of the galaxy number density.
However, correct estimates of the~$\cB$ value (and~$\cA$)
require fully relativistic treatments, as perturbations like velocity and
potential are gauge dependent quantities. 
Furthermore, while Eq.~(\ref{eq:fullGR}) can be derived with the minimal
assumption that the spacetime is described by a perturbed FLRW metric and
photons follow geodesics, the coefficients~$\cB$ and~$\cA$ in 
Eq.~(\ref{eq:full}) are obtained by utilizing the Einstein equations.
Therefore, measuring~$\cB$ and~$\cA$ is equivalent to a direct 
measurement of the relativistic contributions, {\it not} through their
influence on the matter density fluctuation, and
we collectively refer to the coefficients~$\cB$ and~$\cA$
as the general relativistic effects in the galaxy power spectrum. 

\begin{figure}[t]
\centerline{\psfig{file=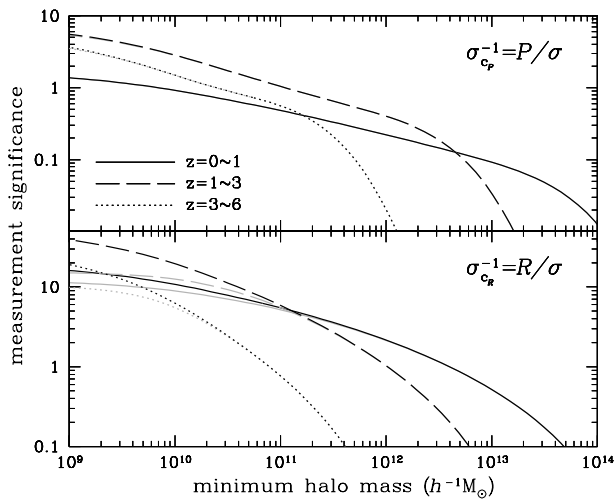, width=3.3in}}\vspace{-10pt}
\caption{Predicted measurement significance of general relativistic
effects in the galaxy power spectrum. All the halos above a given
minimum mass are utilized to take advantage of the multi-tracer
method \cite{SELJA09,HASEDE11}. Three curves represent 
different survey redshift ranges with corresponding 
volume $V=59$, 410, 600~$(\hgpc)^3$. 
Gray curves show the degradation of the measurement significance, when the
evolution bias~$\ee$ and the luminosity function slope~$p$ are marginalized
over.
For the volume-limited sample with constant comoving number density,
$(\cA,~\cB)=(1.7,~1.6)$, $(2.1,~1.5)$, and $(2.2,~1.5)$ at each redshift.}
\label{fig:f1}
\end{figure}

{\it Multi-tracer shot noise cancelling method.$-$}
We consider multiple galaxy samples
with different bias factors for measuring
the general relativistic effects
in the galaxy power spectrum. Using a vector notation, the observed galaxy
fluctuation fields can be written as
\beeq
\label{eq:ebias}
\bdobs=\left[\bb+f\mu_k^2\bI+{\ccA\boldsymbol{\cA}\over (k/\HH)^2}
-i\mu_k{\ccB\boldsymbol{\cB}\over k/\HH}\right]\dv+\bnn~,
\eneq
where $\bb$, $\bI$, $\bnn$ are the linear bias, the multi-dimensional 
identity, and the residual-noise field vectors. 
Similarly, the coefficients~$\cB$ and~$\cA$
become vectors with different values of~$\ee$ and~$p$ for each sample. We
introduced two new parameters~$\ccB$ and~$\ccA$ to generalize
the measurement significance of the coefficients~$\cB$ and~$\cA$ to the case
of multiple galaxy samples --- they are $\ccB=\ccA=1$ in general 
relativity, and measurements of these two parameters amount to the
measurement significance of the two vectors~$\boldsymbol{\cB}$ 
and~$\boldsymbol{\cA}$.
By definition the noise-field is independent of the
matter fluctuation $\AVE{\bnn\dv}=0$, and the square bracket in 
Eq.~(\ref{eq:ebias}) defines the effective bias vector~$\BB$.

In order to assess our ability to measure the general relativistic effects
in the galaxy power spectrum, we employ the Fisher information matrix:
\beeq
\label{eq:fisher}
F_{ij}(k,\mu_k)={\mA\over1+\mA}\up{Re}\left(\mC_{ij}\right)+
{\up{Re}\left(\mB_i\mB_j-\mA\mB_i\mBS_j\right)\over(1+\mA)^2}~, 
\eneq
where $\mA=\BT\EEI\BB\Pv$, $\mB_i=\BT\EEI\BB_i\Pv$, 
$\mC_{ij}=\BT_i\EEI\BB_j\Pv$, 
the shot noise matrix $\EE=\AVE{\bnn\bnn^\up{T}}$, 
the matter power spectrum in the comoving gauge is $\Pv(k)$,  
the parameters $\theta_i=\ccB,\ccA,\bee,\bpp$,
and $\BB_i$ is the derivative of $\BB$ with respect
to the parameters~$\theta_i$.
This formula is a straightforward extension of the Fisher matrix derived in
\cite{HASEDE11}, with the effective bias vector~$\BB$ being a complex vector.
The imaginary part arises solely from the $\cB$-term in Eq.~(\ref{eq:ebias})
and its derivative.

{\it Measurement significance.$-$} For definiteness, we consider full sky
surveys with three different redshift ranges and adopt a set of
cosmological parameters consistent with the WMAP7 results \cite{KOSMET11}.
Given the survey volume~$V$, 
the Fisher matrix in Eq.~(\ref{eq:fisher})
is summed over the Fourier volume, 
where $\kmin=2\pi/V^{1/3}$ and  $\kmax=0.03\hmpci$.
To model the Fisher matrix parameters $\mA$, $\mB_i$, $\mC_{ij}$, we adopt the
halo model description in \cite{HASEET10,HASEDE11}; 
It has been well tested against
a suite of $N$-body simulations with Gaussian and non-Gaussian 
initial conditions. We assume that the galaxy samples have a constant comoving
number density ($\bee=3\bI$) and they are selected in a 
volume-limited survey to eliminate the magnification bias (or $\bpp=0.4\bI$
in practice), but we marginalize over~$\ee$ and~$p$ in 
constraining~$\ccB$ and~$\ccA$ with priors $\sigma_\ee=0.1$
and $\sigma_p=0.05$ \cite{COEIET08}.

Figure~\ref{fig:f1} shows the predicted measurement significance
of the general relativistic effects. In our most optimistic scenario,
the velocity term~$\ccB$ can be measured at more than
10-$\sigma$ confidence level at $z\leq1$, while the gravitational
potential~$\ccA$ is at 1-$\sigma$ significance. 
At higher redshift, though the 
increase in the survey volume is partially cancelled by the lower 
abundance of halos at a fixed mass, a substantial improvement can be achieved 
by going beyond $z=1$. However, as the gray curves show, the uncertainties 
in~$\ee$ and~$p$ need to be controlled beyond the current observational
limit, before further improvement can be realized.

Compared to the estimates $S/N\lesssim1$ obtained by using the standard
power spectrum analysis \cite{YOO10}, the multi-tracer method dramatically 
enhances the measurement significance. Furthermore, a method of
measuring the imaginary part in 
the galaxy power spectrum of two tracers \cite{MCDON09} is fully
implemented in our complex covariance matrix as off-diagonal terms. The
result in \cite{MCDON09} would correspond to 
$\cB/\sigma\simeq1.8$ at $z\leq1$. 
Finally, any degeneracy with cosmological
parameters in measuring the relativistic effects is largely eliminated,
since the underlying matter distribution is cancelled \cite{SELJA09}.

\begin{figure}[t]
\centerline{\psfig{file=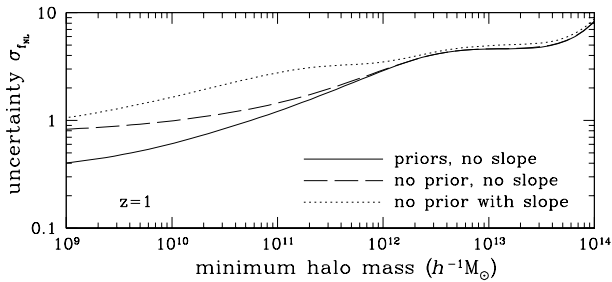, width=3.3in}}\vspace{-10pt}
\caption{Predicted constraints on the primordial non-Gaussianity $\fnl$
from galaxy power spectrum measurements. To facilitate the comparison,
the constraints on $\fnl$ are obtained by using the same survey specifications 
as in \cite{HASEDE11}: $V\simeq50~(\hgpc)^3$ centered at $z=1$. Various
curves show $\sigma_{\fnl}$ with different priors on $\bee$ and $\bpp$.}
\label{fig:f2}
\end{figure}

{\it Primordial non-Gaussianity.$-$} We extend our formalism 
to the primordial non-Gaussian signature in galaxy bias \cite{DADOET08} and
introduce additional parameter~$\fnl$.
Here we only consider the simplest local form of primordial non-Gaussianity
to demonstrate how it can be implemented in the general relativistic 
description, and ignore scale-independent and scale-dependent corrections
(see, e.g., \cite{SLHIET08DESE10DEJESC11a}).

The primordial non-Gaussian signature in galaxy bias can be readily 
implemented in our full general relativistic description with the Gaussian
bias factor in Eq.~(\ref{eq:fullGR}) replaced by
\beeq
b\rightarrow b+3\fnl(b-1)\delta_c~{\OM(z)\HH^2\over T_\varphi(k,z)~k^2}~,
\label{eq:fnl}
\eneq
or equivalently by adding additional term
$3\fnl(b-1)\delta_c\OM(z)/T_\varphi(k,z)$ to the coefficient $\cA$ in 
Eq.~(\ref{eq:full}), where $T_\varphi$ is the transfer function of the
curvature perturbation (see also \cite{BRCRET11,BASEET11,JESCHI11}).

In obtaining the constraint $\sigma_{\fnl}$ on primordial non-Gaussianity,
we set $\ccB=\ccA=1$ and marginalize over~$\ee$ and~$p$. With the
current uncertainties in~$\ee$ and~$p$, the constraint $\sigma_{\fnl}$
(solid in Fig.~\ref{fig:f2}) is nearly identical
to the unmarginalized constraint. 
The dashed curve shows that even with no priors 
on~$\ee$ and~$p$, $\sigma_{\fnl}$ is not inflated
except in the regime with $\sigma_{\fnl}\lesssim2$, because $\cB$
and $\cA$ are affected simultaneously by~$\ee$ and~$p$
but only $\cA$ by $\fnl$. Furthermore, the unique dependence
of $\fnl$ on $b-1$ and $T_\varphi$ in Eq.~(\ref{eq:fnl}) provides the 
multi-tracer method with more leverage to separate it from the general
relativistic effect.

Finally, we allow~$\bee$ and~$\bpp$ to vary as a function of mass
with two logarithmic slope parameters $\alpha_\ee$ and $\alpha_p$:
$\bee=\ee_0\bI+\alpha_\ee\ln(\boldsymbol{M}/M_0)$,
$\bpp=p_0\bI+\alpha_p\ln(\boldsymbol{M}/M_0)$ with $\ee_0=3$, $p_0=0.4$,
$\alpha_\ee=\alpha_p=0$, and $M_0=10^{12}\msun$.
The effects of~$\alpha_e$ and~$\alpha_p$ (dotted in Fig.~\ref{fig:f2})
are sufficiently different from that of~$\fnl$, and
$\sigma_{\fnl}$ asymptotically reaches the floor set by the uncertainties
in~$\ee$ and~$p$.
This demonstrates that the general relativistic effects in the
galaxy power spectrum are {\it not} degenerate with the primordial non-Gaussian
signature. However, if $\fnl$ were to be constrained below unity,
similar precision needs to be achieved in predicting~$\cB$ and~$\cA$.
This requirement can be relaxed by using independent mass estimates,
as $p$ drops out in Eq.~(\ref{eq:full}), further separating $\ee$
from $\fnl$.

{\it Discussion.$-$}
\label{sec:discussion} We have demonstrated that in a typical galaxy survey
at low redshift general relativistic effects in the galaxy power spectrum
can be measured at very high significance by using the multi-tracer method,
providing a unique opportunity to test general relativity on horizon scales.
We also show how the primordial non-Gaussian effect in galaxy bias can
be implemented in the full general relativistic description, and quantitatively
proved that the ability to detect primordial non-Gaussianity is
{\it little} compromised by the presence of general relativistic effects.

Proper application of our method to observations will require a few key
investigations
 beyond our simple treatment considered in this {\it Letter}. First,
while we treated multiple galaxy samples as halos in multiple mass bins,
it is shown \cite{HASEDE11} that a large scatter in the 
mass-observable relation, i.e.,  $\sigma_{\ln M}=0.5$, would degrade the
signal-to-noise ratio by less than a factor of two.
Second, our prediction is based on the halo model description of the shot 
noise matrix, which is tested only for halos at $M\geq10^{12}\msun$, and our
prediction at $M\leq10^{12}\msun$ is an extrapolation. 

We acknowledge useful discussions with Daniel Eisenstein and Pat McDonald.
J.Y. is supported by the SNF Ambizione Grant.
This work is supported by the Swiss National Foundation under contract
200021-116696/1 and WCU grant R32-10130.
M.~Z. is supported by the David and Lucile Packard, the Alfred~P. Sloan,
and the John~D. and Catherine~T. MacArthur Foundations.

\vfill
\bibliography{detect.bbl}

\begin{thebibliography}{24}
\expandafter\ifx\csname natexlab\endcsname\relax\def\natexlab#1{#1}\fi
\expandafter\ifx\csname bibnamefont\endcsname\relax
  \def\bibnamefont#1{#1}\fi
\expandafter\ifx\csname bibfnamefont\endcsname\relax
  \def\bibfnamefont#1{#1}\fi
\expandafter\ifx\csname citenamefont\endcsname\relax
  \def\citenamefont#1{#1}\fi
\expandafter\ifx\csname url\endcsname\relax
  \def\url#1{\texttt{#1}}\fi
\expandafter\ifx\csname urlprefix\endcsname\relax\def\urlprefix{URL }\fi
\providecommand{\bibinfo}[2]{#2}
\providecommand{\eprint}[2][]{\url{#2}}

\bibitem[{\citenamefont{{Yoo} et~al.}(2009)\citenamefont{{Yoo}, {Fitzpatrick},
  and {Zaldarriaga}}}]{YOFIZA09}
\bibinfo{author}{\bibfnamefont{J.}~\bibnamefont{{Yoo}}},
  \bibinfo{author}{\bibfnamefont{A.~L.} \bibnamefont{{Fitzpatrick}}},
  \bibnamefont{and}
  \bibinfo{author}{\bibfnamefont{M.}~\bibnamefont{{Zaldarriaga}}},
  \bibinfo{journal}{\prd} \textbf{\bibinfo{volume}{80}},
  \bibinfo{pages}{083514} (\bibinfo{year}{2009}).

\bibitem[{\citenamefont{{Yoo}}(2010)}]{YOO10}
\bibinfo{author}{\bibfnamefont{J.}~\bibnamefont{{Yoo}}},
  \bibinfo{journal}{\prd} \textbf{\bibinfo{volume}{82}},
  \bibinfo{pages}{083508} (\bibinfo{year}{2010}).

\bibitem[{\citenamefont{{Seljak}}(2009)}]{SELJA09}
\bibinfo{author}{\bibfnamefont{U.}~\bibnamefont{{Seljak}}},
  \bibinfo{journal}{\prl} \textbf{\bibinfo{volume}{102}},
  \bibinfo{pages}{021302} (\bibinfo{year}{2009}).

\bibitem[{\citenamefont{{McDonald}}(2009)}]{MCDON09}
\bibinfo{author}{\bibfnamefont{P.}~\bibnamefont{{McDonald}}},
  \bibinfo{journal}{\jcap} \textbf{\bibinfo{volume}{11}}, \bibinfo{pages}{26}
  (\bibinfo{year}{2009}).

\bibitem[{\citenamefont{{Seljak} et~al.}(2009)\citenamefont{{Seljak}, {Hamaus},
  and {Desjacques}}}]{SEHADE09}
\bibinfo{author}{\bibfnamefont{U.}~\bibnamefont{{Seljak}}},
  \bibinfo{author}{\bibfnamefont{N.}~\bibnamefont{{Hamaus}}}, \bibnamefont{and}
  \bibinfo{author}{\bibfnamefont{V.}~\bibnamefont{{Desjacques}}},
  \bibinfo{journal}{\prl} \textbf{\bibinfo{volume}{103}},
  \bibinfo{pages}{091303} (\bibinfo{year}{2009}).

\bibitem[{\citenamefont{{Hamaus} et~al.}(2011)\citenamefont{{Hamaus}, {Seljak},
  and {Desjacques}}}]{HASEDE11}
\bibinfo{author}{\bibfnamefont{N.}~\bibnamefont{{Hamaus}}},
  \bibinfo{author}{\bibfnamefont{U.}~\bibnamefont{{Seljak}}}, \bibnamefont{and}
  \bibinfo{author}{\bibfnamefont{V.}~\bibnamefont{{Desjacques}}},
  \eprint{arXiv:1104.2321}.

\bibitem[{\citenamefont{{Hamaus} et~al.}(2010)\citenamefont{{Hamaus}, {Seljak},
  {Desjacques}, {Smith}, and {Baldauf}}}]{HASEET10}
\bibinfo{author}{\bibfnamefont{N.}~\bibnamefont{{Hamaus}}},
  \bibinfo{author}{\bibfnamefont{U.}~\bibnamefont{{Seljak}}},
  \bibinfo{author}{\bibfnamefont{V.}~\bibnamefont{{Desjacques}}},
  \bibinfo{author}{\bibfnamefont{R.~E.} \bibnamefont{{Smith}}},
  \bibnamefont{and}
  \bibinfo{author}{\bibfnamefont{T.}~\bibnamefont{{Baldauf}}},
  \bibinfo{journal}{\prd} \textbf{\bibinfo{volume}{82}},
  \bibinfo{pages}{043515} (\bibinfo{year}{2010}).

\bibitem[{\citenamefont{{Bonvin} and {Durrer}}(2011)}]{BODU11CHLE11}
\bibinfo{author}{\bibfnamefont{C.}~\bibnamefont{{Bonvin}}} \bibnamefont{and}
  \bibinfo{author}{\bibfnamefont{R.}~\bibnamefont{{Durrer}}},
  \eprint{arXiv:1105.5280}.
\bibinfo{author}{\bibfnamefont{A.}~\bibnamefont{{Challinor}}} \bibnamefont{and}
  \bibinfo{author}{\bibfnamefont{A.}~\bibnamefont{{Lewis}}},
  \eprint{arXiv:1105.5292}.

\bibitem[{\citenamefont{{Jeong} et~al.}(2011)}]{JESCHI11}
\bibinfo{author}{\bibfnamefont{D.}~\bibnamefont{{Jeong}}},
  \bibinfo{author}{\bibfnamefont{F.}~\bibnamefont{{Schmidt}}},
  \bibnamefont{and} \bibinfo{author}{\bibfnamefont{C.~M.}
  \bibnamefont{{Hirata}}}, \eprint{arXiv:1107.5427}.

\bibitem[]{FOOTNOTE}
\bibinfo{author}{\bibfnamefont{{On 
large scales a more general stochastic relation
between the galaxy number density and the matter density also
reduces to the local form we adopted here [11]. As opposed
to some confusion in literature, this biasing scheme is independent of whether
galaxies are observed.}}}

\bibitem[{\citenamefont{{Scherrer} and {Weinberg}}(1998)}]{SCWE98}
\bibinfo{author}{\bibfnamefont{R.~J.} \bibnamefont{{Scherrer}}}
  \bibnamefont{and} \bibinfo{author}{\bibfnamefont{D.~H.}
  \bibnamefont{{Weinberg}}}, \bibinfo{journal}{\apj}
  \textbf{\bibinfo{volume}{504}}, \bibinfo{pages}{607} (\bibinfo{year}{1998}).

\bibitem[{\citenamefont{{Hwang} and {Noh}}(1999)}]{HWNO99WASL09}
\bibinfo{author}{\bibfnamefont{J.-C.} \bibnamefont{{Hwang}}} \bibnamefont{and}
  \bibinfo{author}{\bibfnamefont{H.}~\bibnamefont{{Noh}}},
  \bibinfo{journal}{\prd} \textbf{\bibinfo{volume}{59}},
  \bibinfo{pages}{067302} (\bibinfo{year}{1999});
\bibinfo{author}{\bibfnamefont{D.}~\bibnamefont{{Wands}}} \bibnamefont{and}
  \bibinfo{author}{\bibfnamefont{A.}~\bibnamefont{{Slosar}}},
  \bibinfo{journal}{\prd} \textbf{\bibinfo{volume}{79}},
  \bibinfo{pages}{123507} (\bibinfo{year}{2009}).

\bibitem[{\citenamefont{{Yoo}}(2009)}]{YOO09}
\bibinfo{author}{\bibfnamefont{J.}~\bibnamefont{{Yoo}}},
  \bibinfo{journal}{\prd} \textbf{\bibinfo{volume}{79}},
  \bibinfo{pages}{023517} (\bibinfo{year}{2009}).

\bibitem[{\citenamefont{{Baldauf} et~al.}(2011)\citenamefont{{Baldauf},
  {Seljak}, {Senatore}, and {Zaldarriaga}}}]{BASEET11}
\bibinfo{author}{\bibfnamefont{T.}~\bibnamefont{{Baldauf}}},
  \bibinfo{author}{\bibfnamefont{U.}~\bibnamefont{{Seljak}}},
  \bibinfo{author}{\bibfnamefont{L.}~\bibnamefont{{Senatore}}},
  \bibnamefont{and}
  \bibinfo{author}{\bibfnamefont{M.}~\bibnamefont{{Zaldarriaga}}},
  \eprint{arXiv:1106.5507}.

\bibitem[{\citenamefont{{Bruni} et~al.}(2011)\citenamefont{{Bruni},
  {Crittenden}, {Koyama}, {Maartens}, {Pitrou}, and {Wands}}}]{BRCRET11}
\bibinfo{author}{\bibfnamefont{M.}~\bibnamefont{{Bruni}}},
  \bibnamefont{et~al.},
  \bibinfo{journal}{arXiv:1106.3999}.

\bibitem[{\citenamefont{{Hui} et~al.}(2008)\citenamefont{{Hui},
  {Gazta{\~n}aga}, and {Loverde}}}]{HUGALO08}
\bibinfo{author}{\bibfnamefont{L.}~\bibnamefont{{Hui}}},
  \bibinfo{author}{\bibfnamefont{E.}~\bibnamefont{{Gazta{\~n}aga}}},
  \bibnamefont{and}
  \bibinfo{author}{\bibfnamefont{M.}~\bibnamefont{{Loverde}}},
  \bibinfo{journal}{\prd} \textbf{\bibinfo{volume}{77}},
  \bibinfo{pages}{063526} (\bibinfo{year}{2008}).


\bibitem[{\citenamefont{{Komatsu} et~al.}(2011)\citenamefont{{Komatsu},
  {Smith}, {Dunkley}, {Bennett}, {Gold}, {Hinshaw}, {Jarosik}, {Larson},
  {Nolta}, {Page} et~al.}}]{KOSMET11}
\bibinfo{author}{\bibfnamefont{E.}~\bibnamefont{{Komatsu}}},
  \bibnamefont{et~al.}, \bibinfo{journal}{\apjs}
  \textbf{\bibinfo{volume}{192}}, \bibinfo{pages}{18} (\bibinfo{year}{2011}).

\bibitem[{\citenamefont{{Cool} et~al.}(2008)\citenamefont{{Cool}, {Eisenstein},
  {Fan}, {Fukugita}, {Jiang}, {Maraston}, {Meiksin}, {Schneider}, and
  {Wake}}}]{COEIET08}
\bibinfo{author}{\bibfnamefont{R.~J.} \bibnamefont{{Cool}}},
  \bibnamefont{et~al.}, \bibinfo{journal}{\apj}
  \textbf{\bibinfo{volume}{682}}, \bibinfo{pages}{919} (\bibinfo{year}{2008}).

\bibitem[{\citenamefont{{Dalal} et~al.}(2008)\citenamefont{{Dalal}, {Dor{\'e}},
  {Huterer}, and {Shirokov}}}]{DADOET08}
\bibinfo{author}{\bibfnamefont{N.}~\bibnamefont{{Dalal}}},
  \bibinfo{author}{\bibfnamefont{O.}~\bibnamefont{{Dor{\'e}}}},
  \bibinfo{author}{\bibfnamefont{D.}~\bibnamefont{{Huterer}}},
  \bibnamefont{and}
  \bibinfo{author}{\bibfnamefont{A.}~\bibnamefont{{Shirokov}}},
  \bibinfo{journal}{\prd} \textbf{\bibinfo{volume}{77}},
  \bibinfo{pages}{123514} (\bibinfo{year}{2008}).

\bibitem[{\citenamefont{{Slosar} et~al.}(2008)\citenamefont{{Slosar}, {Hirata},
  {Seljak}, {Ho}, and {Padmanabhan}}}]{SLHIET08DESE10DEJESC11a}
\bibinfo{author}{\bibfnamefont{A.}~\bibnamefont{{Slosar}}},
  \bibnamefont{et~al.},
  \bibinfo{journal}{\jcap} \textbf{\bibinfo{volume}{8}}, \bibinfo{pages}{31}
  (\bibinfo{year}{2008});

\bibinfo{author}{\bibfnamefont{V.}~\bibnamefont{{Desjacques}}}
  \bibnamefont{and} \bibinfo{author}{\bibfnamefont{U.}~\bibnamefont{{Seljak}}},
  \bibinfo{journal}{\cqg} \textbf{\bibinfo{volume}{27}},
  \bibinfo{pages}{124011} (\bibinfo{year}{2010});

\bibinfo{author}{\bibfnamefont{V.}~\bibnamefont{{Desjacques}}},
  \bibinfo{author}{\bibfnamefont{D.}~\bibnamefont{{Jeong}}}, \bibnamefont{and}
  \bibinfo{author}{\bibfnamefont{F.}~\bibnamefont{{Schmidt}}},
  \eprint{arXiv:1105.3628}.

\end{thebibliography}

\end{document}